\documentclass[twocolumn,prl,preprintnumbers,tightenlines,footinbib,superscriptaddress]{revtex4}

\pdfoutput=1
\usepackage[english]{babel}
\usepackage{amsmath,amssymb,amsfonts}
\usepackage{graphicx}
\usepackage{booktabs}
\usepackage{hyperref}
\usepackage{color}
\usepackage[sort&compress]{natbib}
\usepackage{tikz}
\usepackage{cleveref}

\begin{document}

\title{Some Exact Results in QCD-like Theories}

\author{Hitoshi Murayama}
\email{hitoshi@berkeley.edu, hitoshi.murayama@ipmu.jp, Hamamatsu Professor}
\affiliation{Department of Physics, University of California, Berkeley, CA 94720, USA}
\affiliation{Kavli Institute for the Physics and Mathematics of the
  Universe (WPI), University of Tokyo,
  Kashiwa 277-8583, Japan}
\affiliation{Ernest Orlando Lawrence Berkeley National Laboratory, Berkeley, CA 94720, USA}

\begin{abstract} 
I propose a controlled approximation to QCD-like theories with massless quarks by employing supersymmetric QCD perturbed by anomaly-mediated supersymmetry breaking. They have identical massless particle contents. Thanks to the ultraviolet-insensitivity of anomaly mediation, dynamics can be worked out exactly when $m \ll \Lambda$, where $m$ is the size of supersymmetry breaking and $\Lambda$ the dynamical scale of the gauge theory. I demonstrate that chiral symmetry is dynamically broken for $N_{f} \leq \frac{3}{2} N_{c}$ while the theories lead to non-trivial infrared fixed points for larger number of flavors. While there may be a phase transition as $m$ is increased beyond $\Lambda$, qualitative agreements with expectations in QCD are encouraging and suggest that two limits $m \ll \Lambda$ and $m \gg \Lambda$ may be in the same universality class.
\end{abstract}

\maketitle

\section{Introduction}

It has been a long-standing goal in theoretical physics to understand strongly correlated systems. Well-known examples are high-temperature superconductors in condensed matter physics and strong interaction in particle physics. The latter is described by Quantum ChromoDynamics (QCD) based on $SU(3)$ gauge theory with quarks and gluons as fundamental degrees of freedom. At high energies, its dynamics can be studied with perturbation theory thanks to asymptotic freedom. On the other hand, at low energies, theory becomes strongly coupled at a scale $\Lambda$ that makes fundamental degrees of freedom trapped inside bound states, and typical coupling constants among bound states are $\sim O(4\pi)$ beyond the reach of perturbation theory. In particular, QCD-like theories with massless quarks are believed to dynamically break the chiral symmetry \cite{Nambu:1961tp,Nambu:1961fr} which has been difficult to demonstrated analytically.

Certain strongly coupled theories have exact solutions in the infrared (IR). The important example for the discussion here is supersymmetric QCD (SQCD) whose dynamics was worked out by Seiberg \cite{Seiberg:1994bz,Seiberg:1994pq}. Yet SQCD without the quark mass terms has extra massless degrees freedom, namely squarks and gauginos, which are not present in QCD. No limit of SQCD seems to mimic the dynamics of QCD.

One can introduce supersymmetry breaking to SQCD to mimic non-supersymmetric QCD \cite{Evans:1996hi,Konishi:1996iz,Aharony:1995zh,Evans:1995ia,Evans:1995rv,Evans:1997dz}. In many cases, there is no full theoretical control, or a sign that there is a phase transition when supersymmetry breaking becomes as large as the dynamical scale. If $U(1)_{R}$ is gauged, its $D$-term can make the squarks massive \cite{Pomarol:1999ie}. Its impact is controlled by the $R$-charge even for composite fields and is under a good control. However, gauginos are still massless and the theory does not belong to the same universality class as QCD. Instead, I can introduce an $F$-component to the gauge coupling
\begin{align}
	S = \frac{8\pi^{2}}{g^{2}} + i \vartheta 
	- \theta^{2} \frac{16\pi^{2}}{g^{2}}m_{\lambda} , 
\end{align}
where $\vartheta$ is the vacuum angle, and $m_{\lambda}$ the gaugino mass. All holomorphic quantities are a function of $S$ and the supersymmetry breaking effects are well constrained, while non-holomorphic corrections in K\"ahler potential cannot be controlled, such as mass of composite states, including their signs that are crucial for the symmetry breaking pattern. The papers \cite{Evans:1996hi,Evans:1995ia,Evans:1995rv,Evans:1997dz} avoided this issue by adding larger quark mass to keep the total mass-squared positive, which in turn makes it impossible to study the massless quarks.  Either approach is not suitable to study the low-energy limits of massless QCD.  There are works using string theory \cite{Sakai:2004cn,Sakai:2005yt}, or that using compactification on the abelian subgroup of the chiral symmety \cite{Cherman:2016hcd}, yet the progress has been very limited in the four-dimensional field theories. 

I point out that the anomaly-mediated supersymmetry breaking (AMSB) \cite{Randall:1998uk,Giudice:1998xp} is an exception. It has a property called ultraviolet (UV) insensitivity, and the supersymmetry breaking effects are completely controlled at all energy scales with any description. Therefore, the known exact results in SQCD can be generalized to include the effects of AMSB \footnote{I used this argument in Ref.~\cite{Harnik:2003rs} to work out phenomenology of the ``fat Higgs'' model but only for the case of quantum modified moduli space where the impact of AMSB is the least important.}. In addition, SQCD with AMSB (ASQCD) makes both squarks and gauginos massive, and hence its massless particle content is identical to that of QCD. Therefore, ASQCD is a QCD-like theory. As the supersymmetry breaking scale $m$ is increased beyond $\Lambda$, the ASQCD reduces completely to QCD. If ASQCD and QCD are continuously connected, the low-energy limits of QCD must be in the same universality class as that of ASQCD. There is no guarantee, however, that there is no phase transition at a critical value $m_{c} \sim O(\Lambda)$, and two theories are not continuously connected. Yet the results I find in this Letter are encouraging and the two limits $m \ll \Lambda$ and $m \gg \Lambda$ may well belong to the same universality class.

\section{Anomaly Mediation}

Anomaly mediation of supersymmetry breaking (AMSB) can be formulated with the Weyl compensator $\Phi = 1 + \theta^{2} m$ \cite{Pomarol:1999ie} that appears in the supersymmetric Lagrangian as
\begin{align}
	{\cal L} &= \int d^{4} \theta \Phi^{*} \Phi K + \int d^{2} \theta \Phi^{3} W + c.c.
\end{align}
Here, $K$ ($W$) is the K\"ahler potential (superpotential) of the theory, and $m$ is the parameter of supersymmetry breaking. When the theory is conformal, $\Phi$ can be removed from the theory by rescaling the fields $\phi_{i} \rightarrow \Phi^{-1} \phi_{i}$. On the other hand, violation of conformal invariance leads to supersymmetry breaking effects. Solving for auxiliary fields, the superpotential leads to the tree-level supersymmetry breaking terms
\begin{align}
	{\cal L}_{\rm tree} &= m \left( \phi_{i} \frac{\partial W}{\partial \phi_{i}} - 3 W \right)
	+ c.c.
	\label{eq:tree}
\end{align}
Dimensionless coupling constants do not lead to supersymmetry breaking effects because of the conformal invariance at the tree-level. However, conformal invariance is anomalously broken due to the running of coupling constants, and there are loop-level supersymmetry breaking effects in tri-linear couplings, scalar masses, and gaugino masses,
\begin{align}
	A_{ijk} (\mu) &= - \frac{1}{2} (\gamma_{i} + \gamma_{j} + \gamma_{k})(\mu) m, 
	\label{eq:A} \\
	m_{i}^{2}(\mu) &= - \frac{1}{4} \dot{\gamma}_{i}(\mu) m^{2}, 
	\label{eq:m2} \\
	m_{\lambda}(\mu) &= - \frac{\beta(g^{2})}{2g^{2}}(\mu) m. \label{eq:mlambda}
\end{align}
Here, $\gamma_{i} = \mu\frac{d}{d\mu} \ln Z_{i}(\mu)$, $\dot{\gamma} = \mu \frac{d}{d\mu} \gamma_{i}$, and $\beta(g^{2}) = \mu \frac{d}{d\mu} g^{2}$. In general, physical masses are the sum of contributions from the superpotential (tree-level or non-perturbative), tree-level AMSB \eqref{eq:tree} and loop-level AMSB (\ref{eq:A},\ref{eq:m2},\ref{eq:mlambda}).

In the supersymmetric $SU(N_{c})$ QCD with $N_{f}$ flavors in the fundamental representation (SQCD), I find at the leading order of ASQCD
\begin{align}
	m_{Q}^{2} = m_{\tilde{Q}}^{2} &= \frac{g^{4}}{(8\pi^{2})^{2}} 2 C_{i} (3N_{c} - N_{f}) 
	m^{2}, 
	\label{eq:mQ2} \\
	m_{\lambda} &= \frac{g^{2}}{16\pi^{2}}  (3 N_{c} - N_{f}) m .
\end{align}
Here, $C_{i} = \frac{N_{c}^{2}-1}{2N_{c}}$, and $m_{{Q}}^{2} > 0$ in the range $N_{f} < 3N_{c}$ where the theory is asymptotically free. Therefore, the squarks and gauginos acquire mass and the massless particle content is identical to that of non-supersymmetric QCD. As $m$ is increased beyond the dynamical scale of the gauge theory $\Lambda$, gluinos and squarks can be integrated out, and the theory does become QCD. I do not know a priori whether the change in $m$ is continuous. There may or may not be a phase transition as $m$ crosses $O(\Lambda)$. Nonetheless, SQCD with AMSB is continuously connected to QCD, and I hope to learn something about the dynamics of QCD by studying SQCD with AMSB.

The most remarkable property of the anomaly mediated supersymmetry breaking is its ultraviolet-insensitivity. The expressions for the supersymmetry breaking parameters above depend on wave function renormalization and running coupling constants, which jump when heavy fields are integrated out from the theory. It turns out that the threshold corrections from the loops of heavy fields precisely give the necessary jump. Therefore the above expressions remain true at {\it all}\/ energy scales and depend only on the particle content and interactions present at that energy scale. This point can be verified explicitly in perturbative calculations, and is very transparent in the $\overline{\rm DR}$ scheme \cite{Boyda:2001nh}. 

One way to intuitively understand the ultraviolet-insensitivy is the analogy to quantum field theory in curved spacetime.  To describe QCD in a curved spacetime, I couple the QCD Lagrangian to the background spacetime metric. When QCD confines, I switch to the chiral Lagrangian, and I couple it to the same metric. This is because the back reaction of QCD dynamics to the metric is suppressed by the Planck scale and can be safely ignored. For anomaly-mediated supersymmetry breaking, the Weyl compensator can be viewed as a part of the background supergravity multiplet. Ignoring the back reaction to the superspacetime, I couple the field theory to the same supergravity background no matter what non-perturbative dynamics takes place. 

Since the low-energy dynamics of SQCD is well understood thanks to Seiberg, I couple its low-energy limit to AMSB to work out the ground state exactly. In particular, I am interested in the symmetry of the ground state exactly in the limit $m \ll \Lambda$ under a full theoretical control. I will then discuss how it may be connected to the dynamics of QCD as $m$ is increased beyond $\Lambda$ hoping there is no phase transition, so that the ASQCD and QCD belong to the same universality class. What I find below is encouraging. I assume $N_{c} \geq 3$ in the discussions below.



\section{$N_{f} < N_{c}$}

\begin{figure}[t]
	\includegraphics[width=0.7\columnwidth]{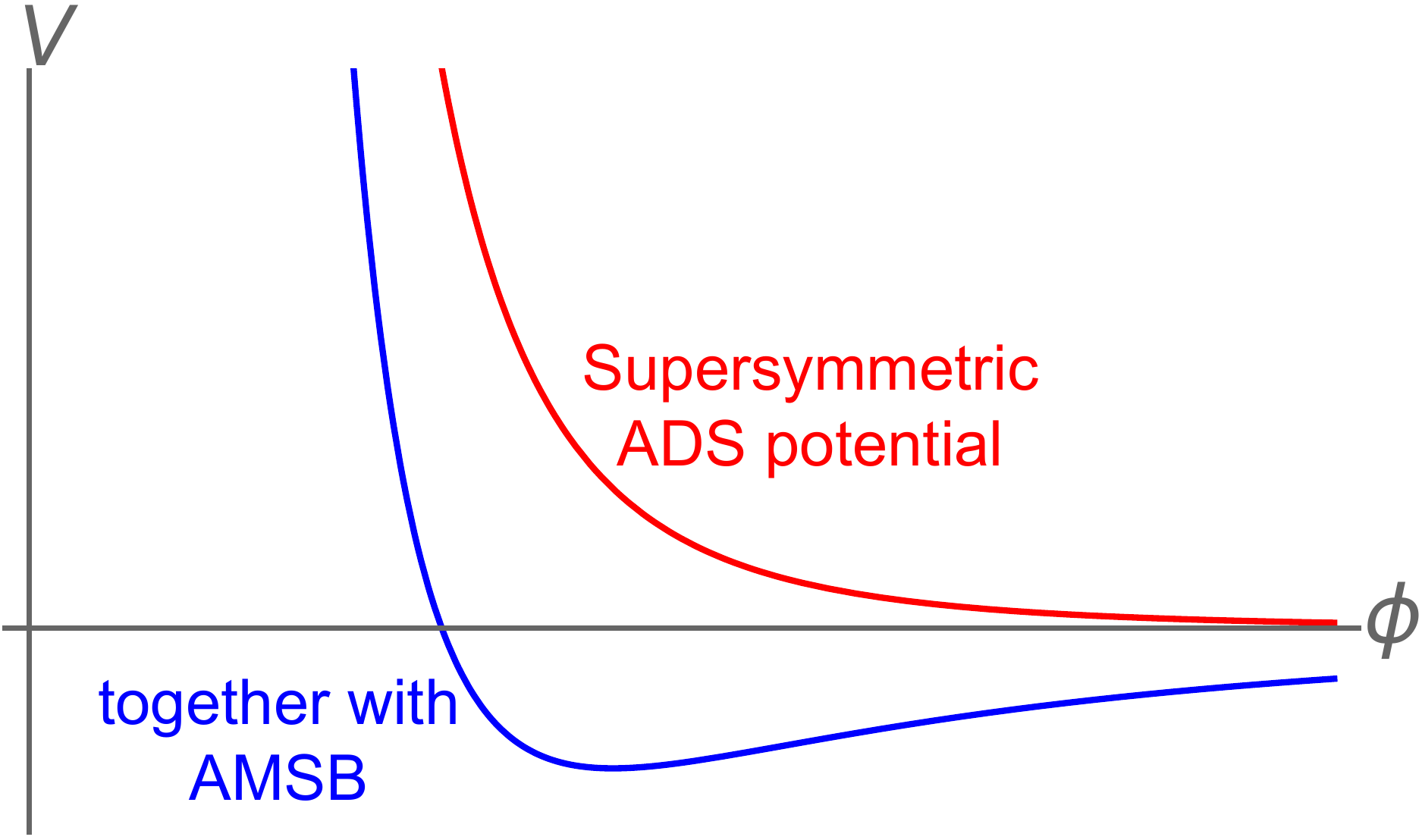}
	\caption{Schematics of the potential for $N_{f} < N_{c}$. The red curve is for SQCD with run-away behavior, while the blue curve for ASQCD has a well-defined minimum. }
	\label{fig:ADS}
\end{figure}

The dynamics is described in terms of the meson fields $M^{ij}$ with the non-perturbative Affleck--Dine--Seiberg (ADS) superpotential
\begin{align}
	W = (N_{c} - N_{f}) \left( \frac{\Lambda^{3N_{c}-N_{f}}}{{\rm det} M} \right)^{1/(N_{c}-N_{f})}
	\ .
\end{align}
The SQCD has a run-away potential and hence no ground states. When $M \gg \Lambda^{2}$, $M_{ij} = M \delta_{ij}$ describes the $D$-flat direction
\begin{align}
	Q = \tilde{Q} = \left(\begin{array}{ccc} 
		1 & \cdots & 0 \\ 
		\vdots &\ddots & \vdots \\
		0 & \cdots & 1 \\ \hline
		0 & \cdots & 0 \\
		\vdots & \vdots & \vdots \\
		0 & \cdots & 0
		\end{array} \right) \phi, \qquad M = \phi^{2}		.
\end{align}
The upper part is an $N_{f} \times N_{f}$ block, while the lower part $(N_{c}-N_{f}) \times N_{f}$.
Therefore the Lagrangian along this direction in ASQCD is
\begin{align}
	{\cal L} & = \int d^{4} \theta \Phi^{*} \Phi 2N_{f} \phi^{*} \phi \nonumber \\
	&+ \int d^{2} \theta \Phi^{3} (N_{c}-N_{f}) 
	\left( \frac{\Lambda^{3N_{c}-N_{f}}}{\phi^{2N_{f}}} \right)^{1/(N_{c}-N_{f})}	\ .
\end{align}
The corresponding potential is
\begin{align}
	V &=\left|2N_{f} \frac{1}{\phi} 
	\left(\frac{\Lambda^{3N_{c}-N_{f}}}{\phi^{2N_{f}}} \right)^{1/(N_{c}-N_{f})} \right|^{2}
	\nonumber \\
	& - (3N_{c} - N_{f} )
	m \left( \frac{\Lambda^{3N_{c}-N_{f}}}{\phi^{2N_{f}}} \right)^{1/(N_{c}-N_{f})}
	+ c.c.
\end{align}
Note that there is now a well-defined minimum (see Fig.~\ref{fig:ADS}),
\begin{align}
	M_{ij} &= \Lambda^{2} 
	\left( \frac{4N_{f}(N_{c}+N_{f})}{3N_{c}-N_{f}} \frac{\Lambda}{m} \right)^{(N_{c}-N_{f})/N_{c}}
	\delta_{ij} .
\end{align}
The minimum is indeed at $M_{ij} \gg \Lambda^{2}$ which justifies the weakly-coupled analysis. The mass for mesons from AMSB is loop suppressed and hence can be ignored. The $SU(N_{f})_{Q} \times SU(N_{f})_{\tilde{Q}}$ flavor symmetry is dynamically broken to $SU(N_{f})_{V}$. The massless particle spectrum is the corresponding Nambu--Goldstone bosons (pions) \footnote{ The case $N_{f}=1$ is special as there is no non-anomalous flavor symmetry and hence the spectrum is gapped.}. The scalar and fermion partners of the Nambu--Goldstone bosons (NGBs) have mass that grows with $m$. Naively increasing $m$ beyond $\Lambda$, the only remaining degrees of freedom are massless NGBs. This seems to match the expectations in QCD with small number of flavors. There is no sign of a phase transition and the two limits are likely continuously connected.

\section{$N_{f} = N_{c}$}

This is the case of quantum modified moduli space described by the superpotential
\begin{align}
	W = X ({\rm det} M - \tilde{B} B- \Lambda^{2N_{c}}).
\end{align}
Given the successful and highly non-trivial anomaly matching conditions, it is believed that the K\"ahler potential is regular at the origin for meson and baryon superfields. By going to canonical normalization of the fields, I find the superpotential
\begin{align}
	W = X \left(\lambda \frac{{\rm det} M}{\Lambda^{N_{c}-2}} 
	- \kappa \tilde{B} B- \Lambda^{2}\right).
\end{align}
Here, $\lambda, \kappa$ are dimensionless coupling constants. I find two candidate ground states which I work out to the first order in $m \ll \Lambda$. 

One is
\begin{align}
	M^{ij} = \lambda^{-1/N_{c}} \Lambda \delta^{ij}, &\qquad
	B = \tilde{B} = 0, \nonumber \\
	X = \lambda^{-2/N_{c}} m, &\qquad
	V = -N_{c} \lambda^{-2/N_{c}} m^{2} \Lambda^{2} . \label{eq:V1}
\end{align}
The massless spectrum is the NGBs of $SU(N_{f})_{Q} \times SU(N_{f})_{\tilde{Q}} / SU(N_{f})_{V}$. The anomalies are matched by the Wess--Zumino term \cite{Wess:1971yu,Witten:1983tw} induced by integrating out massive mesinos. 

The other is
\begin{align}
	M^{ij} = 0 , &\qquad
	B = \tilde{B} = \kappa^{-1/2} \Lambda, \nonumber \\
	X = \kappa^{-1} m, &\qquad
	V = - (2 \kappa)^{-1} m^{2} \Lambda^{2} . \label{eq:V2}
\end{align}
The massless spectrum is the NGB of spontaneously broken $U(1)_{B}$ and mesinos that match the anomalies of $SU(N_{f})_{Q} \times SU(N_{f})_{\tilde{Q}}$.

I cannot determine which minimum is lower without knowing $\lambda$ and $\kappa$. However, the first one is likely be continuously connected to QCD, while it is difficult to imagine there are massless mesinos in the non-supersymmetric limit. Here I rely on the naive dimensional analysis \cite{Luty:1997fk,Cohen:1997rt} which suggests $\lambda \approx (4\pi)^{N_{c}/2}$ and $\kappa \approx 4\pi$. Then I find the first minimum Eq.~\eqref{eq:V1} is $V \approx - N_{c} \frac{1}{4\pi} m^{2} \Lambda^{2}$, while the second Eq.~\eqref{eq:V2} is $V \approx - \frac{1}{8\pi} m^{2} \Lambda^{2}$. Therefore, Eq.~\eqref{eq:V1} is the global minimum, where we find massless NGBs of $SU(N_{f})_{Q} \times SU(N_{f})_{\tilde{Q}} / SU(N_{f})_{V}$ with $f_{\pi} \approx \Lambda/(4\pi)^{2}$, as well as baryons that acquire mass $m_{B} \approx m$. This is an analytic demonstration that QCD with 3 colors and 3 flavors break chiral symmetry with massless pions and massive baryons.


\section{$N_{f} = N_{c}+1$}

\begin{figure}[t]
	\includegraphics[width=0.7\columnwidth]{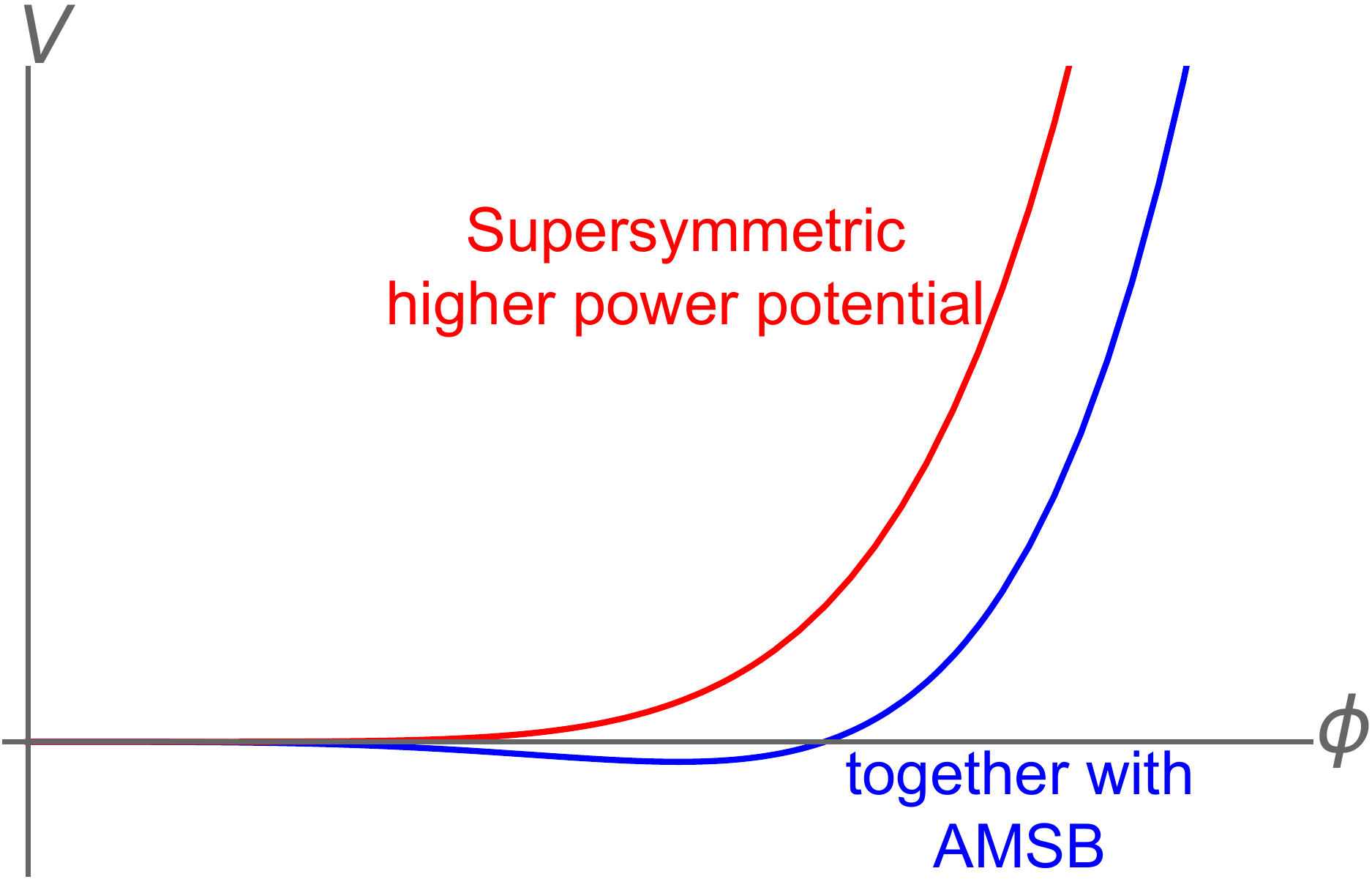}
	\caption{Schematics of the potential for $s$-confining and free magnetic phases. The red curve is for SQCD with power-law behavior, while the blue curve for ASQCD has a well-defined minimum. }
	\label{fig:irrelevant}
\end{figure}

The theory is $s$-confining with the dynamical superpotential
\begin{align}
	W = \frac{1}{\Lambda^{2N_{f}-3}} \left( {\rm det} M - \tilde{B}_{i} M^{ij} B_{j} \right).
\end{align}
Again given the successful and highly non-trivial anomaly matching conditions, it is believed that the K\"ahler potential is regular at the origin for meson and baryon superfields. By going to canonical normalization of the fields, I find the superpotential
\begin{align}
	W = \lambda\frac{{\rm det} M}{\Lambda^{N_{f}-3}}  - \kappa\tilde{B} M B .
\end{align}
The corresponding potential for ASQCD for the direction $M^{ij} = \phi \delta^{ij}$ is
\begin{align}
	V &= N_{f} \lambda^{2} \frac{\left| \phi \right|^{2N_{f}-2}}{\Lambda^{2N_{f}-6}}
	- \lambda (N_{f}-3) m \phi^{N_{f}} + c.c.
\end{align}
It has a minimum at (see Fig.~\ref{fig:irrelevant})
\begin{align}
	\phi &=\Lambda\left( \frac{(N_{f}-3)m}{(N_{f}-1)\lambda\Lambda} \right)^{1/(N_{f}-2)} 
	\ll \Lambda,
\end{align}
justifying the analysis with the canonical K\"ahler potential. The mass for mesons from AMSB is loop suppressed and hence can be ignored. The vacuum energy is $V \sim -O(m^{(2N_{f}-2)} \Lambda^{(2N_{f}-6)})^{1/(N_{f}-2)}$. The massless spectrum is once again the NGBs of $SU(N_{f})_{Q} \times SU(N_{f})_{\tilde{Q}} / SU(N_{f})_{V}$. There is no sign of phase transition as $m$ is increased beyond $\Lambda$.

Note that ${\rm rank}\ M = N_{c}$ classically. The vacuum with ${\rm rank}\ M = N_{f} > N_{c}$ here and below is a genuinely strong-coupled effect in the electric theory. Yet the description with the meson field in the magnetic theory is weakly coupled because of its IR-free dynamics and its small expectation value, and is hence justified.

Note that the term $\kappa \tilde{B} M B$ is renormalizable and the corresponding tri-linear supersymmetry breaking is loop suppressed $A_{\kappa} \sim m \kappa^{2}/16\pi^{2}$. 
This leads to another candidate ground state with $M_{11} = B^{1} = \tilde{B}^{1} \sim m / 16\pi^{2}\kappa$ with the vacuum energy $V \sim - O(\frac{m}{16\pi^{2}})^{4}$. Clearly the other minimum is deeper. 

\section{$N_{c}+2 \leq N_{f} \leq \frac{3}{2} N_{c}$}

This range is in the free magnetic phase described by an $SU(N_{f}-N_{c})$ SQCD as the low-energy theory below the dynamical scale $\Lambda$ with the superpotential 
\begin{align}
	W = \kappa \tilde{q}_{i} M^{ij} q_{j}\ .
\end{align}
Here, $q$ ($\tilde{q}$) are dual (anti-)quarks. The mesons $M$ are canonically normalized. Similarly to the $s$-confining case of $N_{f}=N_{c}+1$, there is one-loop tri-linear coupling that allows for a minimum with
\begin{align}
	q = \tilde{q} = \left(\begin{array}{ccc} 
		1 & \cdots & 0 \\ 
		\vdots &\ddots & \vdots \\
		0 & \cdots & 1 \\ \hline
		0 & \cdots & 0 \\
		\vdots & \vdots & \vdots \\
		0 & \cdots & 0
		\end{array} \right) \phi, &\quad
	M = \left( \begin{array}{ccc|ccc}
		1 & \cdots & 0 & 0 & \cdots & 0 \\
		\vdots & \ddots & \vdots & \vdots & \ddots & \vdots \\
		0 & \cdots & 1 & 0 & \cdots & 0 \\ \hline
		0 & \cdots & 0 & 0 & \cdots & 0 \\
		\vdots & \ddots & \vdots & \vdots & \ddots & \vdots \\
		0 & \cdots & 0 & 0 & \cdots & 0 
		\end{array} \right) \chi
\end{align}
Here, the upper block is $(N_{f}-N_{c}) \times (N_{f}-N_{c})$ while the lower block $N_{c} \times (N_{f}-N_{c})$. This direction has the vacuum energy $V \sim - O(\frac{m}{16\pi^{2}})^{4}$.

There is, however, a deeper minimum. It is found along the direction $M^{ij} = \phi \delta^{ij}$. All dual quarks are massive and can be integrated out, yielding a pure $SU(N_{f}-N_{c})$ supersymmetric Yang--Mills theory that leads to a gaugino condensate. The effective superpotential is \cite{Intriligator:2006dd}
\begin{align}
	W = (N_{f}-N_{c}) 
	\left( \frac{\kappa^{N_{f}}{\rm det} M }{\Lambda^{3N_{c}-2N_{f}}} \right)^{1/(N_{f}-N_{c})} \ .
	\label{eq:ISS}
\end{align}
Note that it is a higher-dimension operator and hence an irrelevant operator that does not modify the IR-free nature of the mesons. The potential in ASQCD is
\begin{align}
	V &= N_{f}\Lambda^{4} \left|\frac{\kappa\phi}{\Lambda}\right|^{2N_{c}/(N_{f}-N_{c})} 
	\nonumber \\
	& - (2N_{f}-3N_{c}) m \Lambda^{3} \left( \frac{\kappa\phi}{\Lambda} \right)^{N_{f}/(N_{f}-N_{c})}
	+ c.c.
\end{align}
The minimum is at (see Fig.~\ref{fig:irrelevant})
\begin{align}
	\phi &= \kappa^{-1}\Lambda 
	\left( \frac{2N_{f}-3N_{c}}{N_{c}} \frac{m}{\Lambda} \right)^{(N_{f}-N_{c})/(2N_{c}-N_{f})}
	\ll \Lambda
\end{align}
which justifies the analysis with the canonical K\"ahler potential for $\phi$. The mass for mesons from AMSB is loop suppressed and hence can be ignored. The massless spectrum is once again the NGBs of $SU(N_{f})_{Q} \times SU(N_{f})_{\tilde{Q}} / SU(N_{f})_{V}$. There is no sign of phase transition as $m$ is increased beyond $\Lambda$.

\section{$\frac{3}{2} (N_{c} + 1) < N_{f} < 3(N_{c}+1)$}

The dual theory is the same as the free magnetic case. However, the superpotential Eq.~\eqref{eq:ISS} becomes relevant and does not justify the analysis with the canonical K\"ahler potential for mesons.

Going back to SQCD, this is the case where both electric and magnetic theories flow to IR fixed points. The conformal dimension of superfields is given by $D = \frac{3}{2}R$ where $R$ is the non-anomalous $U(1)_{R}$ charge for unitary representation of superconformal algebra. The wave function renormalization factors are \cite{deGouvea:1998ft} 
\begin{align}
	Z_{Q}(\mu) = Z_{\tilde{Q}}(\mu) 
	&= \left( \frac{\mu}{\mu'} \right)^{(3N_{c}-N_{f})/N_{f}}
	Z_{Q,\tilde{Q}} (\mu'), \\
	Z_{M}(\mu) & = \left( \frac{\mu}{\mu'} \right)^{(6N_{c}-4N_{f})/N_{f}} 
	Z_{M}(\mu'), \\
	Z_{q}(\mu) = Z_{\tilde{q}} (\mu) 
	&= \left( \frac{\mu}{\mu'} \right)^{(2N_{f}-3N_{c})/N_{f}} 
	Z_{q,\tilde{q}} (\mu') .
\end{align}
Here, $\mu < \mu' < \Lambda_{*}$, and $\Lambda_{*}$ is the energy scale where the theory reaches the IR fixed point. I find
\begin{align}
	\gamma_{Q} = \frac{3N_{c}-N_{f}}{N_{f}}, \quad
	\gamma_{M} = \frac{6N_{c}-4N_{f}}{N_{f}}, \quad
	\gamma_{q} = \frac{2N_{f}-3N_{c}}{N_{f}}.
\end{align}
Since $\dot{\gamma}_{i} = 0$, all scalar masses vanish. Also, $A_{\kappa} \propto \gamma_{M} + 2 \gamma_{q} = 0$. Finally, the exact NSVZ beta-function \cite{Novikov:1983uc,ArkaniHamed:1997mj} is
\begin{align}
	\beta(g^{2}) &= g^{4} \frac{-(3N_{c} - N_{f} ) + N_{f} \gamma_{Q}}{8\pi^{2} - N_{c} g^{2}}
	= 0
\end{align}
in the electric theory and
\begin{align}
	\beta(g^{2}) &= g^{4} 
	\frac{-(2N_{f} - 3N_{c} ) + N_{f} \gamma_{q}}{8\pi^{2} - (N_{f}-N_{c}) g^{2}}
	= 0
\end{align}
in the magnetic theory. Therefore the gaugino masses vanish as well. Namely, the IR limit of the theory has no impact of AMSB because of its conformal invariance. 

The discussion above assumes the theory has already reached the IR fixed point $\mu <  \Lambda_{*}$. In general, the approach to IR fixed points is not logarithmic but rather a power law, because
\begin{align}
	\frac{d g^{2}}{d t} &= \beta(g^{2}) = \beta'_{*} (g^{2} -g_{*}^{2}) + O(g^{2} -g_{*}^{2})^{2},
\end{align}
and hence (for $\mu' > \mu > \Lambda_{*}$)
\begin{align}
	\frac{g^{2} (\mu) -g_{*}^{2}}{g^{2} (\mu') -g_{*}^{2}}
	&=  \left( \frac{\mu}{\mu'} \right)^{\beta'_{*}}.
\end{align}
Therefore the gaugino mass near the fixed point is
\begin{align}
	m_{\lambda}(\mu) &= -m \beta'_{*} \frac{g^{2} (\mu') -g_{*}^{2}}{2g^{2}(\mu)}
	\left( \frac{\mu}{\mu'} \right)^{\beta'_{*}}
\end{align}
and power suppressed. Namely the approach to IR fixed point is in general rather quick.

What I find here is that non-supersymmetric theories become supersymmetric in the IR, an {\it emergent supersymmetry}\/! 

What happens when $m$ is increased? Since the IR theory does not {\it know}\/ AMSB, it most likely stays as superconformal theory up to $m \sim \Lambda$. Once $m$ crosses $O(\Lambda)$, it may exhibit a phase transition. In QCD with a large number of flavors, it is believed that the theory flows to IR fixed points. The best evidence is the Banks--Zaks fixed point for large $N_{c}$ and $N_{f}$ \cite{Banks:1981nn}. Therefore, there may be a phase transition from a superconformal field theory to a non-supersymmetric conformal field theory. It may also be the case that the emergent superconformal symmetry persists even when $m > \Lambda$. Right now I cannot determine which is the case, but both possibilities are truly fascinating.

\section{$N_{f} \geq 3 N_{c}$}

In this case, the theory is IR-free, and the squark mass Eq.~\eqref{eq:mQ2} is tachyonic.  There is no stable ground state, and hence ASQCD is not continuously connected to QCD.

\section{$Sp(N_{c})$}

The above analysis easily carries over to SQCD based on $Sp(N_{c})$ gauge groups \cite{Intriligator:1995ne}. I do not repeat the discussions here, but point out that $SU(2N_{f})$ flavor symmetry is dynamically broken to $Sp(N_{f})$ for $N_{f} \leq \frac{3}{2}(N_{c}+1)$ in ASQCD. For $\frac{3}{2}(N_{c}+1) < N_{f} < 3(N_{c}+1)$, ASQCD flows to superconformal field theory. This again matches expectation in non-supersymmetric QCD based on $Sp(N_{c})$ gauge groups \cite{Witten:1983tx}. The case $N_{c}=1$ is special as the dynamical superpotential for $N_{f}=3$ is renormalizable. Therfefore the mass for mesons from AMSB cannot be ignored. An analysis with loop-level AMSB is warranted and will be discussed elsewhere.

\section{$SO(N_{c})$}

The dynamics of SQCD based on $SO(N_{c})$ gauge groups are similar, with dynamical symmetry breaking of $SU(N_{f})$ flavor symmetry to $SO(N_{f})$. his case is interesting given there is an unambiguous notion of confinement. There are, however, important differences due to exotic composites and multiple branches \cite{Intriligator:1995id}. I do not attempt to discuss it here, and a dedicated study is warranted \footnote{Csaba Cs\`aki, Andrew Gomes, Hitoshi Murayama, Telem Ofri, {\it in preparation}.}.

\section{Conclusions}

I proposed a new analysis method of dynamics of QCD theories by continuously connecting them to supersymmetric QCD with anomaly-mediated supersymmetry breaking. They share the same massless particle contents, and their symmetries appear to be in the same universality class. This is especially true for small number of flavors. The analysis here is a demonstration of dynamical chiral symmetry breaking in QCD in a controlled approximation. For larger number of flavors, supersymmetric QCD with anomaly-mediated supersymmetry breaking actually restores supersymmetry in the IR ({\it emergent}\/ supersymmetry), leading to superconformal field theories. It supports the conventional wisdom that QCD with a large number of flavors flows to an IR fixed point. I believe this analysis method opens up new avenues to study dynamics of non-supersymmetric gauge theories including chiral ones \cite{Csaki:2021xhi,Csaki:2021aqv}.


\begin{acknowledgments}

I thank Csaba Cs\'aki and Yuji Tachikawa for careful reading of the manuscript and useful comments. This work was supported by the Director, Office of Science, Office of
High Energy Physics of the U.S. Department of Energy under the
Contract No. DE-AC02-05CH11231, by the NSF grant
PHY-1915314, by the JSPS Grant-in-Aid for
Scientific Research JP20K03942, MEXT Grant-in-Aid for Transformative Research Areas (A)
JP20H05850, JP20A203, by World Premier International Research Center Initiative, MEXT, Japan, and Hamamatsu Photonics, K.K.
\end{acknowledgments}

\bibliographystyle{utcaps_mod}
\bibliography{refs}

\end{document}